\documentclass{PoS}
\usepackage{comment}
\usepackage[authoryear]{natbib}
\bibpunct{(}{)}{;}{a}{}{,}

\title{The Astrophysics of Star Formation Across Cosmic Time at $\gtrsim$10\,GHz with the Square Kilometre Array}

\ShortTitle{Astrophysics at $\gtrsim$10\,GHz with the SKA}

\author{Eric J. Murphy$^{1}$, 
Mark T. Sargent$^{2}$\speaker{}, 
Rob J. Beswick$^{3}$,
Clive Dickinson$^{3}$,
Ian Heywood$^{4,5}$,
Leslie K. Hunt$^{6}$,
Minh T. Hyunh$^{7}$, 
Matt Jarvis$^{8,9}$,
Alexander Karim$^{10}$,
Marita Krause$^{11}$,
Isabella Prandoni$^{12}$, 
Nicholas Seymour$^{13}$,
Eva Schinnerer$^{14}$,
Fatemeh S. Tabatabaei$^{14}$, 
Jeff Wagg$^{15}$ 
\\
$^{1}$IPAC, Caltech, MC 220-6, Pasadena CA, 91125, USA;
$^{2}$Astronomy Centre, Department of Physics and Astronomy, University of Sussex, Brighton BN1 9QH, UK; 
$^{3}$Jodrell Bank Centre for Astrophysics, The University of Manchester, Oxford Rd, Manchester M13 9PL, UK; 
$^{4}$CSIRO Astronomy \& Space Science, P.O. Box 76, Epping, NSW 1710, Australia; 
$^{5}$RATT, Department of Physics and Electronics, Rhodes University, PO Box 94, Grahamstown 6140, South Africa; 
$^{6}$INAF - Osservatorio Astrofisico di Arcetri, Largo E. Fermi 5, 50125, Firenze, Italy; 
$^{7}$ICRAR, University of Western Australia, Stirling Highway, Crawley, Western Australia 6009, Australia; 
$^{8}$Astrophysics, University of Oxford, Denys Wilkinson Building, Keble Road, Oxford, OX1 3RH, UK; 
$^{9}$Department of Physics, University of the Western Cape, Bellville 7535, South Africa; 
$^{10}$Argelander-Institut f\"ur Astronomie, Auf dem H\"ugel 71, 53121 Bonn, Germany; 
$^{11}$Max-Planck-Institut f\"ur Radioastronomie, Auf dem H\"ugel 69, 53121 Bonn, Germany; 
$^{12}$INAF - Istituto di Radioastronomia, Via Gobetti 101, Bologna, Italy; 
$^{13}$CASS, P.O. Box 76, Epping, NSW, 1710, Australia; 
$^{14}$Max-Planck-Institut f\"ur Astronomie, K\"onigstuhl 17, 69117 Heidelberg, Germany; 
$^{15}$Square Kilometre Array Organisation, Lower Withington, Cheshire, UK
\\
Email: \email{emurphy@ipac.caltech.edu}}

\abstract{
In this chapter, we highlight a number of science investigations that are enabled by the inclusion of Band~5 ($4.6-13.8$\,GHz) for SKA1-MID science operations, while focusing on the astrophysics of star formation over cosmic time.  
For studying the detailed astrophysics of star formation at high-redshift, surveys at frequencies $\gtrsim$10\,GHz have the distinct advantage over traditional $\sim$1.4\,GHz surveys as they are able to yield higher angular resolution imaging while probing higher rest frame frequencies of galaxies with increasing redshift, where emission of star-forming galaxies becomes dominated by thermal (free-free) radiation.  
In doing so, surveys carried out at $\gtrsim$10\,GHz provide a robust, dust-unbiased measurement of the massive star formation rate by being highly sensitive to the number of ionizing photons that are produced.  
To access this powerful star formation rate diagnostic requires that Band~5 be available for SKA1-MID.  
We additionally present a detailed science case for frequency coverage extending up to 30\,GHz during full SKA2 operations, as this allows for highly diverse science while additionally providing contiguous frequency coverage between the SKA and ALMA, which will likely be the two most powerful interferometers for the coming decades.  
To enable this synergy, it is crucial that the dish design of the SKA be flexible enough to include the possibility of being fit with receivers operating up to 30\,GHz.
}

\FullConference{
Advancing Astrophysics with the Square Kilometre Array\\
June 8-13, 2014\\
Giardini Naxos, Italy}
\vspace*{0pt}

\newcommand{\skipthis}[1]{}
\newcommand\nar{New Astronomy Reviews}
\newcommand\apj{ApJ}
\newcommand\apjl{ApJ}
\newcommand\apjs{ApJ}
\newcommand\aap{A\&A}
\newcommand\aj{AJ}
\newcommand\mnras{MNRAS}
\newcommand\nat{Nature}
\newcommand\araa{ARA\&A}

\newcommand{\farcs}{\mbox{\ensuremath{.\!\!^{\prime\prime}}}}

\begin{document}

\section{Introduction}
\noindent

Currently, the SKA1 System Baseline Design \citep{ska1_baseline} states that SKA1-MID dishes will be capable of operations up to at least 20\,GHz, but only three out of five frequency bands (spanning a total range of $0.350-13.8$\,GHz) will be populated.  
In this chapter, we discuss the scientific motivation for the inclusion of the highest-frequency band (i.e., Band~5; $4.6-13.8$\,GHz) during SKA1-MID operations, as it opens up unique science unachievable with any other band.  
Observations at frequencies $\gtrsim$10\,GHz have the distinct advantage over those at $\sim$1\,GHz as they probe higher rest frame frequencies of galaxies with increasing redshift, where emission becomes dominated by {\it thermal} (free-free) radiation.  
Radio free-free emission is both largely extinction free and can be directly related to the ionizing photon rate arising from newly formed massive stars.  
Evidence for this has been indicated by detailed studies of star-forming regions in the Galaxy \citep[e.g.,][]{pgm67a}, nearby dwarf irregulars \citep[e.g.][]{kg86},  the nuclei of normal galaxies \citep[e.g.,][]{th83,th94} and starbursts \citep[e.g.,][]{kwm88,th85}, Wolf-Rayet galaxies \citep[e.g.,][]{2000AJ....120..244B}, as well as high resolution investigations of super star clusters within nearby blue compact dwarfs \citep[e.g.,][]{thb98, kj99,2009AJ....137.3788J}.  
By {\it directly} measuring the ionizing photon rate from massive stars independent of dust, such observations provide an unbiased estimate of the massive star formation rate.  
With $\sim$200\,km baselines, observations at $\gtrsim$10\,GHz can achieve a maximum angular resolution of $\lesssim$0\farcs03, sampling $\approx$250\,pc scales within disk galaxies at $z\gtrsim1$ and allowing for detailed investigations of galaxy energetics and distributed star formation during the peak of cosmic star formation activity.  

In this chapter, we focus on a number of critical advancements in our understanding of {\it star formation} over cosmic time that can only be realized through observations at $\gtrsim$10\,GHz with the SKA.   
The science goals presented here are largely based on the ultra-deep SKA1-MID/Band~5 reference survey outlined in \citet{prandoniAASKA14}.  
We also focus on galaxies at large cosmological distances, and refer the reader to the chapter by \citet{beswickAASKA14} for a detailed discussion on studying the astrophysics of star formation within nearby galaxies.  
While Band~5 is also critical for studies of AGN, we only provide a few examples of such instances as a comprehensive overview of this topic is beyond the scope of the chapter.  
For example, by being in the Faraday-thin regime, observations at $\gtrsim$10\,GHz allow for a clear view of the intrinsic polarization properties for detailed exploration of the internal physics, magnetic fields, and thermal plasma environments of AGN cores.  
Additionally, observations at $\gtrsim$10\,GHz provide access to maser line emission that can be used to identify AGN.  

We additionally present a detailed science case for frequency coverage extending up to 30\,GHz (where ALMA Band 1 will begin) during full SKA2 operations, as this is critical for a proper accounting of the energetic processes powering galaxies; non-thermal (synchrotron), free-free (bremsstrahlung), and thermal dust emission all start to contribute at somewhat commensurate levels at such frequencies.  
The science that is enabled by the inclusion of this frequency range is highly diverse.  
For instance, this spectral window opens up the possibility for investigations of the star formation law, which relates the star formation rate and gas surfaces densities in galaxies, by providing access to the low-$J$ rotational lines of CO $J=1\rightarrow0$ and $J=2\rightarrow1$ for galaxies in the redshift range of $z=2.8 - 10.5$.  
Such a capability will be highly synergistic with ALMA observations that map out the peak of the dust emission for galaxies in a similar redshift range.    

\begin{figure}[t]
\centering
\begin{center}
\includegraphics[scale=.75]{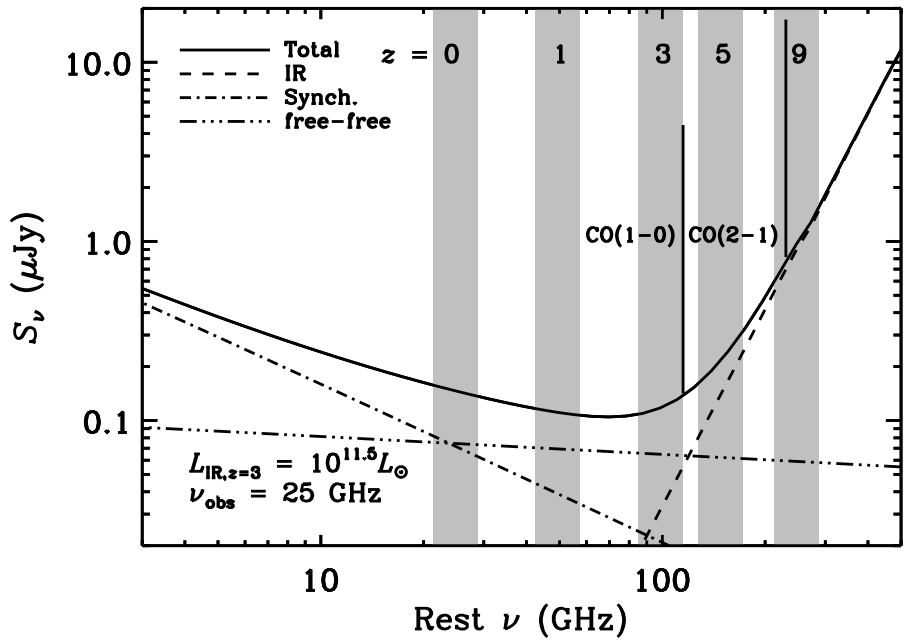}
\caption{\footnotesize 
A rest frame model radio-to-infrared galaxy spectrum having an IR (8-1000$\mu$m) luminosity of $L_{\rm IR} = 10^{11.5}\,L_{\odot}$ at redshift 3.  
The shaded regions indicate where a 25\,GHz observation (see \S\protect \ref{sec:ska}), having 30\% bandwidth in the observed frame, would redshift for $z=0, 1, 3, 5,$ and 9.  
Individual emission components (free-free, synchrotron, and thermal dust) are shown, indicating that at this frequency range between 30 and 100\,GHz (in the rest frame) is most sensitive to free-free emission, making this spectral window extremely robust for accurately measuring the star formation rates of high-redshift galaxies.  
Additionally shown are CO ($J=1\rightarrow0$) and CO ($J=2\rightarrow1$) spectral lines having a line-to-IR flux ratio of M\,82 and 1000\,km\,s$^{-1}$ widths.  
These lines, which are essential for accurately measuring the molecular gas content of galaxies as they trace the bulk of the cold molecular gas component, are accessible in the redshift range of $z=2.8 - 10.5$ over the full $20-30$\,GHz spectral window.  
}
\label{fig:spec}
\end{center}
\end{figure}

\section{Astrophysics Enabled by the Inclusion of Band~5 in SKA1-MID}

%

\subsection{Detailed Radio Spectral Analyses with Band~5}

Synchrotron emission provides unique information on the most powerful phase of star formation (i.e., supernova), as well as the most energetic components of the interstellar medium (i.e., the magnetic fields and cosmic rays). 
The latter each contribute significantly to the total pressure of the interstellar gas, which is important for the onset of star formation, formation of galactic bubbles, and outflows \citep{1966ApJ...145..811P, 2001RvMP...73.1031F, 2005ARAA..43..337C}. 
To what extent cosmic rays and magnetic fields influence the energy balance, and thus the evolution of galaxies, remains an open and highly debated question.  
This, and the origins and energetics of the synchrotron emission in general, can be addressed especially well through detailed studies of the radio spectral energy distribution (SED). 
The fact that at radio wavelengths most surveys have targeted a single radio frequency/band (mainly 1.4 GHz) has prevented a proper radio-SED analysis both on galaxy-wide scales and locally within individual galaxies. 
However, star-formation timescales and histories (SFH) can shape radio spectral indices and lead to larger variations than currently known.  

The combination of the various SKA1-MID reference surveys outlined in the overview chapter \citep{prandoniAASKA14} will provide unprecedented information about the shape of the radio spectrum of low- and high-redshift sources.  
This, in turn, will yield a much better understanding of the relative contributions of thermal and non-thermal emission as the thermal fraction increases with frequency.  
For star-forming galaxies near the apparent peak epoch in the star formation rate density at $z\sim 2-3$, the emission observed in Band 5 at $\sim10$~GHz arises at a rest-frame frequency of $30-40$~GHz, probing the spectral regime where  free-free emission begins to dominate over synchrotron (e.g., see Figure \ref{fig:spec}). 
This means that it is crucial to examine how radio spectra evolve with time and whether they change significantly with SFH \citep[e.g.,][]{2009ApJ...706..482M, schleicher13}.
Young starbursts in the local Universe have flat thermal-like radio spectral indices \citep{deeg93, roussel03, hunt04, hunt05} and high star formation rate surface-density regions are fainter at low radio frequencies than more tenuous regions \citep{2011ApJ...737...67M,heesen14}.
At high redshift, there is currently no consensus on age-dependent spectral indices of starbursts: while some distant starbursts appear to have a flatter radio spectral index \citep{huntmaiolino05,valt11} there is also evidence to the contrary \citep{bourne11,thomson14}.

The unprecedented sensitivity of SKA1-MID and SKA2 will open a wide window to the radio Universe, which is ideal for coherent multi-band surveys allowing for the first time robust radio-SED analyses for distinct galaxy building blocks such as central starbursts/AGNs, spiral arms, star forming regions/complexes and the diffuse interstellar media (see also \S\ref{sec:resolve}).  
SKA1-MID capabilities at $\gtrsim$10\,GHz will be fundamental for assessing truly how the star formation rate density evolves over cosmic time.  
Only with high-frequency observations can the SKA provide necessary constraints for any changes of radio spectral index with redshift, as expected for increased dominance of free-free emission in younger objects \citep{hunt05,hirashita06}. 

There is growing evidence that starbursts deviate from the ``star-forming main sequence" because of intense, spatially concentrated star formation activity, which is reflected by the increased dust temperature
\citep{1991ApJ...378...65C, chanial07, melbourne12, hayward12, symeonidis13, magnelli14} and by a flattening of the radio spectral index \citep{2013ApJ...768....2M}.
Radio spectral indices may also be important to help diagnose mergers \citep{2013ApJ...777...58M}, and estimate their contribution to the starburst fraction at high redshift, currently estimated to be about 10-15\% \citep{rodighiero11, sargent12}. 
Radio-SED analysis with SKA1-MID will address the most pressing questions on: 
(a) the origin of the synchrotron emission in various galactic environments, 
(b) the change in the shape of the SEDs, reflecting changes in radiative processes due to varying physical conditions (e.g., synchrotron spectral index with redshift), and
(c) the balance between the total energy budget of galaxies emitted in radio and infrared over cosmic time. 
Such investigations can only be conducted using Band~5 in combination with lower frequency observations to cover the necessary frequency range for a coherent radio-SED analysis.

\begin{figure}[t]
\center{\hspace*{-1.7cm}
\resizebox{18cm}{!}{
\includegraphics[scale=.43]{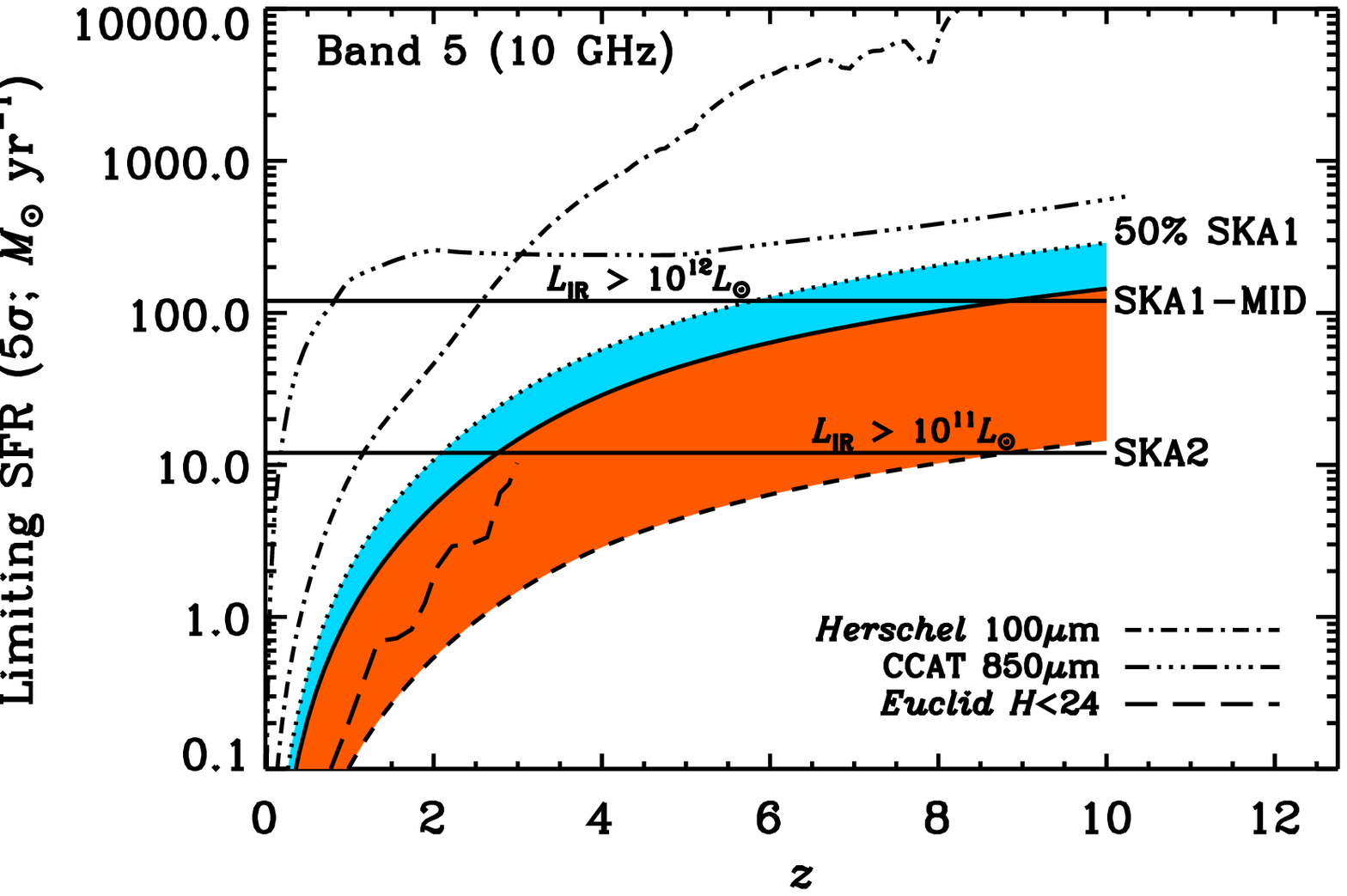}
\includegraphics[scale=.43]{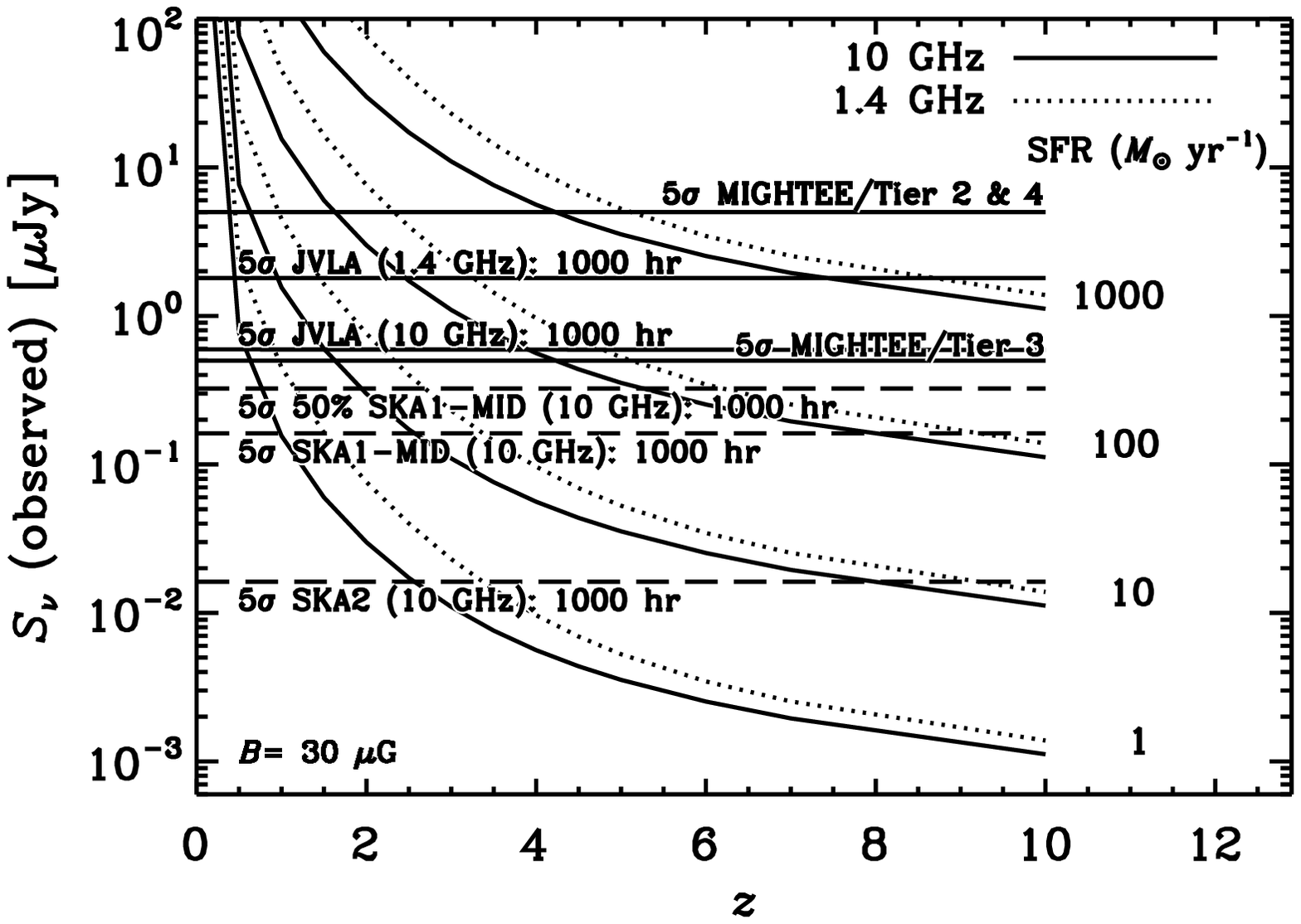}
}
\caption{\footnotesize
{\it Left:} The expected 10\,GHz selection function (5\,$\sigma$) for unresolved star-forming galaxies after a 1000\,hr observation with SKA1-MID and SKA2 given in units of limiting star formation rate as described by the ultra-deep Band~5 reference survey in \citet{prandoniAASKA14}.  
Sensitivities are taken from \citet{ska1_imgperf} assuming 5\,GHz of usable bandwidth and a 0\farcs1 synthesized beam.  
The case of a 50\% less sensitive SKA1-MID is also shown.  
For comparison, we superimpose the (5\,$\sigma$) selection functions for ultra-deep GOODS-S {\it Herschel}100\,$\mu$m data \citep[dot-dashed line;][]{2013AA...553A.132M}, which are the deepest {\it Herschel} far-infrared extragalactic survey data ever taken, the CCAT confusion limit at 850\,$\mu$m (triple dot-dashed line; assuming a 30\,m dish), and for a simulated {\it Euclid} wide survey limited at $H < 24$\,mag \citep[long-dashed line;][]{ciliegiAASKA14}.  
Even with optimistic far-infrared SED assumptions, SKA1-MID surveys should be nearly one and two orders of magnitude more sensitive than 850\,$\mu$m CCAT observations and the deepest space-based far-infrared survey data at $z\gtrsim4$, respectively, providing the most sensitive observations to obscured star formation at high redshifts.  
{\it Right:} The expected 10\,GHz (solid line) and 1.4\,GHz (dotted line) flux densities for galaxies of different star formation rate 
assuming an intrinsic magnetic field strength of 30\,$\mu$G.  
The depths of possible future surveys taken by the JVLA, MeerKAT (MIGHTEE), and SKA are shown.  
Since the non-thermal emission from star-forming galaxies should be suppressed due to increased inverse-Compton scattering of cosmic-ray electrons off of the CMB, the discrepancy between the point source sensitivity requirements of surveys at 1.4 and 10\,GHz falls below a factor of $\lesssim2$ by $z\gtrsim4$.  
\label{fig:senscurv}
}}
\end{figure}

\subsection{Studying Star Formation in the High-Redshift Universe with Band~5}
To date, deep field radio surveys aimed at measuring the star formation history of the Universe have been carried out almost exclusively at 1.4\,GHz, and are thus most sensitive to non-thermal emission processes from galaxies.  
This exclusiveness of low frequency surveys has largely been because of primary beam and sensitivity considerations given the steep spectra ($S_{\nu} \propto \nu^{-0.7}$) of local star-forming galaxies.    
To then construct a picture for the star formation history of the Universe from these low frequency surveys requires an indirect conversion to a star formation rate based on the empirically derived far-infrared/radio correlation \citep{1985ApJ...298L...7H, 2001ApJ...554..803Y}.  
However, as already stated above, surveys at frequencies $\gtrsim$10\,GHz have the distinct advantage of yielding higher angular resolution imaging while probing higher rest frame frequencies of galaxies with increasing redshift, where emission becomes dominated by {\it thermal} (free-free) radiation and {\it directly} provides a dust-unbiased measurement of the instantaneous rate of formation of massive stars.  

The ability to measure free-free radiation directly becomes increasingly important at high redshifts given the increased uncertainties when using standard star formation rate diagnostics.  
Rest-frame FUV observations provide a direct measure of emission from the photospheres of young, massive stars, however it is often hampered by extinction effects that vary spatially within, and among, galaxies.   
Additionally, changes in metallicity and gas-to-dust ratios as a function of redshift may result in galaxies becoming increasingly dust-free at high-$z$, thus making conversions of infrared dust emission to star formation rates difficult to interpret.  

Observations carried out by SKA1-MID using Band~5 at $\approx$10\,GHz will sample the rest frame $\approx$30\,GHz emission from galaxies at $z\sim2$, providing one of the most accurate measurements for star formation activity at high-$z$.  
Results from recent work using the Robert C. Byrd Green Bank Telescope (GBT) and the Karl G. Jansky Very Large Array (JVLA) find that 33\,GHz thermal fractions on $\sim$kpc scales for a range of local galaxy nuclei and extranuclear star-forming regions are $\sim$80\%, on average \citep{2011ApJ...737...67M, 2012ApJ...761...97M}.
Additionally, global investigations of starburst radio spectra (M\,82, NGC\,253, and NGC\,4945) similarly suggest $\gtrsim$80\% thermal fractions at 33\,GHz \citep{2011MNRAS.416L..99P}.  
As we push the detection of star-forming galaxies out to significantly higher redshifts, we anticipate even larger thermal fractions due to increased inverse-Compton losses to cosmic-ray electrons off of the CMB with redshift, whose energy density scales as $U_{\rm CMB} \sim (1+z)^{4}$; 
non-thermal emission from galaxies should become severely suppressed with increasing redshift, making this frequency range ideal for accurate estimates of star formation activity at high $z$, unbiased by dust \citep{2008ApJ...689..883C, 2009ApJ...706..482M}.   

It is worth stressing that, independent of radio spectral index variations, observations at $\gtrsim$10\,GHz are more easily interpreted at high redshift because they become dominated by free-free emission, yielding highly accurate star formation rate estimates even when the non-thermal spectral index is unknown.  
As a quantitative example, even in the absence of any information for the intrinsic non-thermal spectrum, other than the measured distribution for local galaxies \citep[i.e., $ \alpha^{\rm NT} = -0.83$; $ \sigma_{\alpha^{\rm NT}} = 0.13$;][]{nkw97}, by having such a long lever arm over which to measure the radio spectral index using deep 10\,GHz data ensures that a typical uncertainty on the thermal fractions, and thus estimated star formation rates, is {\it only $\sim$30\%} for 20$\sigma$ detections at both 1.4 and 10\,GHz (40\% for 5$\sigma$ detections).    
This uncertainty is half of the dispersion measured for the far-infrared--radio correlation (used to convert 1.4\,GHz data to a star formation rate), and is drastically smaller than the uncertainties associated with extinction correcting rest-frame UV data, as well as the factor of $\sim$2 uncertainty associated with IR luminosity star formation calibrations \citep[e.g.,][]{2012ApJ...761...97M}.  
And, if the thermal fractions do increase as a function of redshift relative to what is measured for local galaxies, this uncertainty will only decrease. 

In the left panel of Figure \ref{fig:senscurv} the 10\,GHz (5\,$\sigma$) selection functions for star-forming galaxies are shown for SKA1-MID and SKA2 after a 1000\,hr exposure as a function of redshift.  
Even with SKA1-MID observations alone, deep surveys will reach an untouched part of parameter space, being able to detect extremely faint galaxies hosting low levels of star formation at extremely high redshifts.  
This is exactly the population of star-forming galaxies for which synchrotron emission should become severely suppressed by CMB effects.  
At such a depth, SKA1-MID observations will be nearly two orders of magnitude more sensitive than the deepest far-infrared (i.e., 100\,$\mu$m) survey carried out with the {\it Herschel} space telescope \citep{2013AA...553A.132M}, as well as approximately an order of magnitude more sensitive than a 30\,m CCAT confusion limited survey at 850\,$\mu$m.  
Furthermore, if SKA1-MID is limited to 50\% of its expected sensitivity, the impact on the Band~5 science described here will be minimal (e.g., see left panel of Figure \ref{fig:senscurv}).  

Unlike observations at 1.4\,GHz \citep[e.g.,][]{2008MNRAS.386.1695S, 2009ApJ...690..610S, 2010ApJS..188..178M} those at 10\,GHz are dominated by thermal emission beyond $z\sim2$ even for large galaxy magnetic field strengths.  
Consequently, star-forming galaxies at increasing redshift should have increasingly flatter spectra due to the suppression of their synchrotron emission by inverse-Compton scattering off the CMB, making higher frequency surveys nearly as sensitive to the same population of star-forming galaxies detected in lower frequency surveys.  
This is illustrated in right panel of Figure \ref{fig:senscurv}, where we plot the expected 1.4 and 10\,GHz flux densities of star-forming galaxies for a range of star formation rates as a function of redshift.  
The ratio of 1.4 to 10\,GHz (observed-frame) flux densities is expected to decrease significantly with increasing redshift.  
Also shown are the anticipated (5\,$\sigma$) 10\,GHz sensitivities for SKA1-MID and SKA2 after a single 1000\,hr exposure compared to JVLA and MeerKAT capabilities.  
After such a deep integration at $\sim$10\,GHz, SKA1-MID and SKA2 should be able to detect galaxies forming starts at $\sim$100 and 10\,$M_{\odot}\,$yr$^{-1}$, at all redshifts, respectively.  

\begin{figure}[t]
\center{\hspace*{-1.7cm}
\resizebox{18cm}{!}{
\includegraphics[scale=.43]{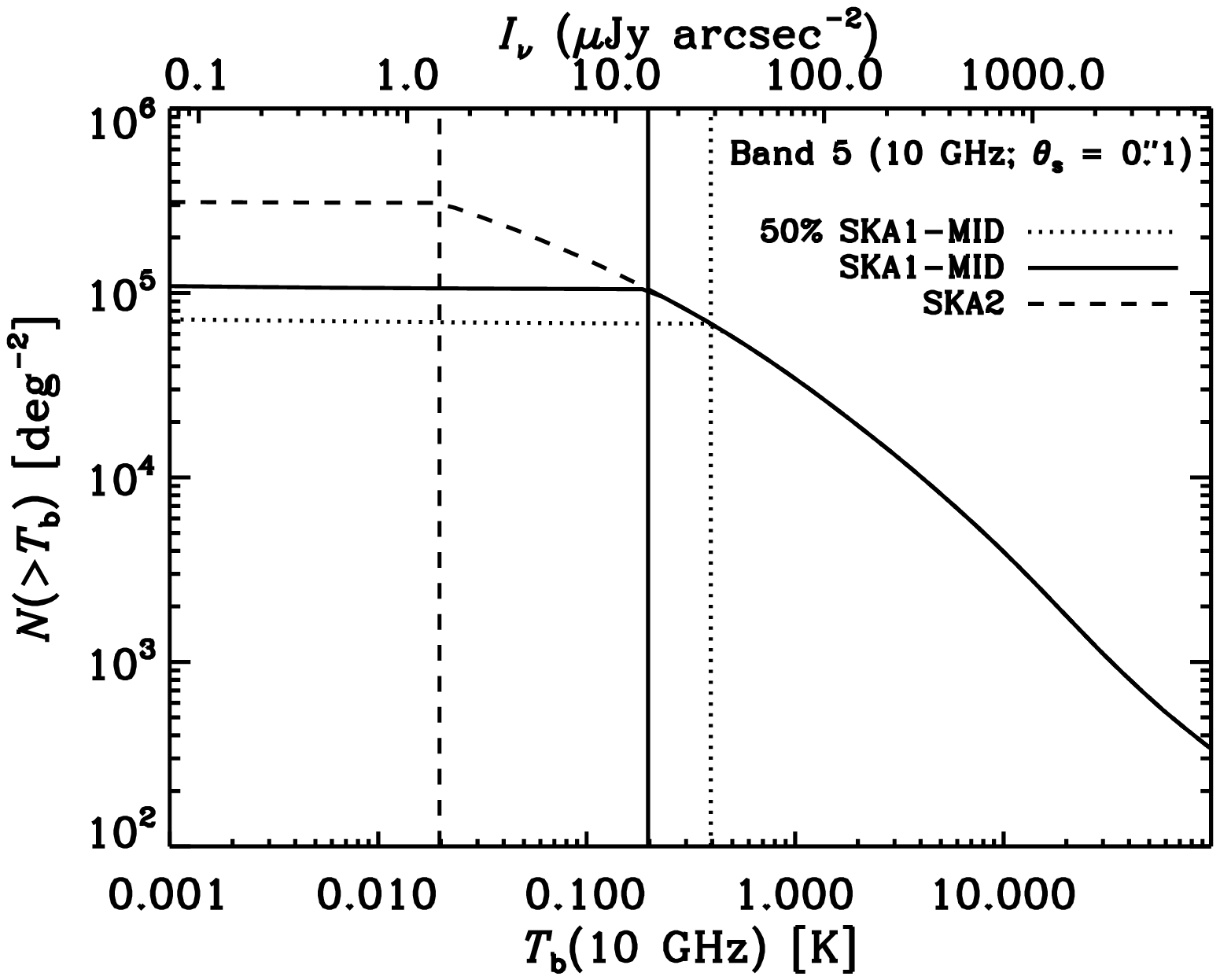}
\includegraphics[scale=.43]{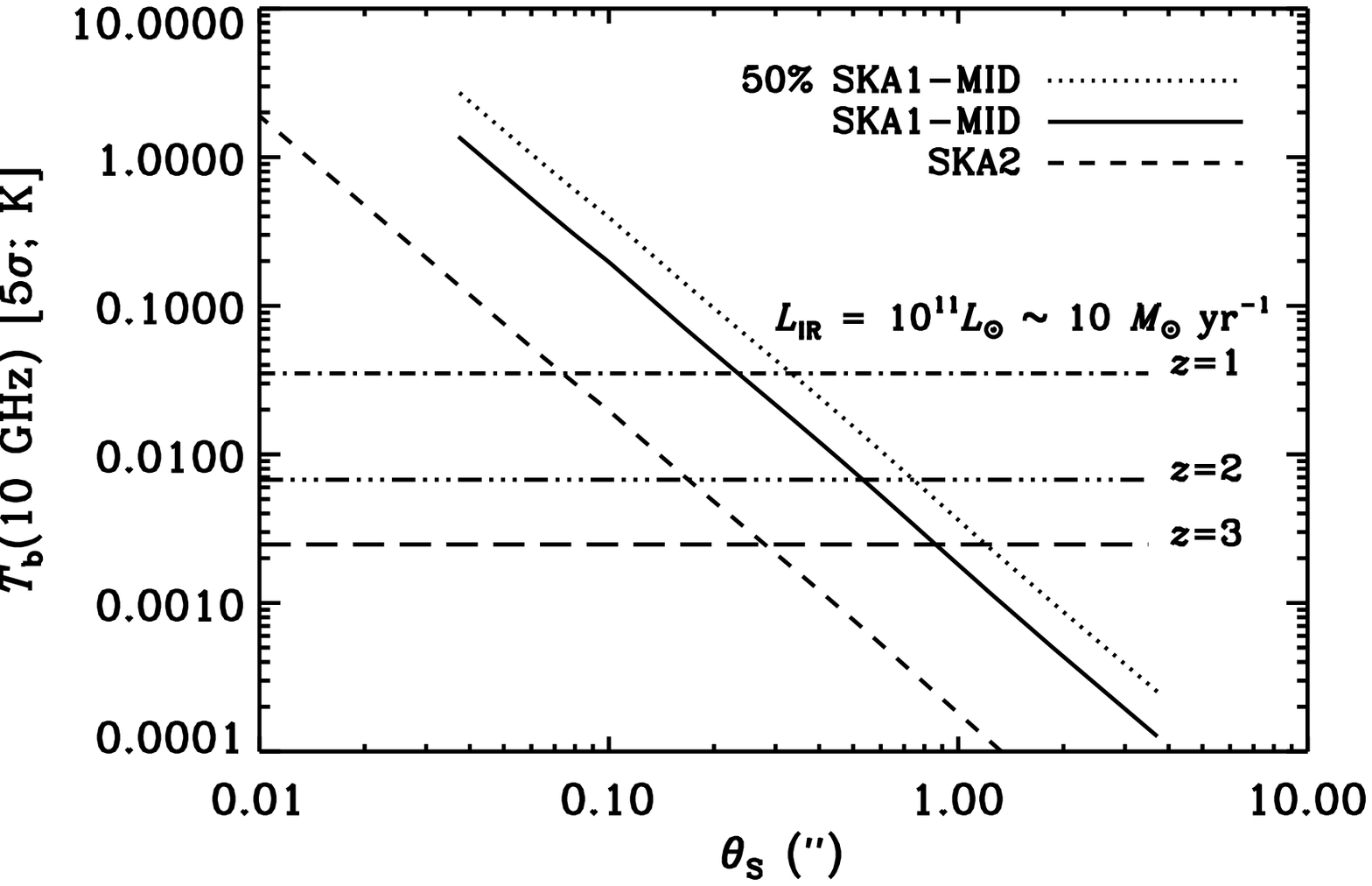}
}
\caption{\footnotesize
{\it Left:} The anticipated cumulative extragalactic source counts for a Band-5 deep field based on the S$^{3}$ simulated radio sky \citep{2008MNRAS.388.1335W}.  The vertical line indicates the 5$\sigma$ surface brightness sensitivity for 1000\,hr observations with SKA1-MID and SKA2.  
Brightness temperature sensitivies are taken from \citet{ska1_imgperf} assuming 5\,GHz of usable bandwidth and a 0\farcs1 synthesized beam.  
The anticipated source density at these surface brightness limits is $\approx$30 and $\approx$85\,arcmin$^{-2}$, respectively.  
The case of a 50\% less sensitive SKA1-MID is also shown, for which the corresponding source density is $\approx$20\,arcmin$^{-2}$.
{\it Right:} The brightness temperature sensitivity (5\,$\sigma$) of the SKA as a function of synthesized beam, assuming uniform sensitivity for baselines up to 200 and 2000\,km baselines for SKA1-MID and SKA2, respectively.  
Additionally plotted is the expected brightness temperature of a $L_{\rm IR} = 10^{11}\,L_{\odot}$ galaxy assuming an intrinsic source size of 0\farcs65 (corresponding to a physical size of $\sim$5\,kpc at $z = 1$, 2, and 3).  
Even with SKA1-MID, observations at 10\,GHz should be able to resolve sources forming stars at $\sim$10\,$M_{\odot}\,$yr$^{-1}$ up to $z\sim2$ under the conservative assumption of uniformly distributed emission.  
\label{fig:tbplots}
}}
\end{figure}

While the new WIDAR correlator on the JVLA is finally allowing users to investigate the high-$z$ Universe with deep-field pointings at these higher frequencies, it is only with the SKA that such observations will yield source counts commensurate with current lower frequency surveys, opening up a completely new parameter space for galaxy evolution studies in the radio.  
This is illustrated in the left panel of Figure \ref{fig:tbplots}, where we show the expected cumulative 10\,GHz source counts for 1000\,hr exposures with SKA1-MID and SKA2 assuming a synthesized beam of $\theta_{\rm s} = 0\farcs1$ based on the S$^{3}$ simulated sky \citep{2008MNRAS.388.1335W}.  
Such an observation corresponds to the ultra-deep SKA1-MID/Band~5 reference survey outlined in \citet{prandoniAASKA14}.  
The anticipated source density at these 5$\sigma$ surface brightness limits is $\approx$30 and $\approx$85\,arcmin$^{-2}$ with SKA1-MID and SKA2, respectively, resulting in a total of $\approx$840 and $\approx$2500 detections per each single pointing deep field, respectively.  
For the case of a 50\% less sensitive SKA1-MID, the corresponding source density is $\approx$20\,arcmin$^{-2}$, resulting in a total of $\approx$550 detections per each single pointing.  
Thus, even though the primary beam at $\gtrsim$1.4\,GHz is $\approx$50 times larger than that at 10\,GHz, deep Band 5 surveys with the SKA will yield statistically significant samples allowing for more complete (e.g., in frequency coverage) investigations on the radio properties of galaxies over cosmic time.

\subsubsection{Resolving Star-Forming Disks During the Peak of the Cosmic Star Formation History \label{sec:resolve}}
With $\sim$200\,km baselines, observations at $\gtrsim$10\,GHz will be able to achieve a maximum angular resolution of $\lesssim$0\farcs03, sampling $\approx$250\,pc scales within disk galaxies at $z\gtrsim1$, providing an extinction-free view for the morphologies of dusty star-bursting galaxies that dominate the star formation activity between $1 \lesssim z \lesssim 3$.  
While probing such fine physical scales requires sources that are extremely bright, the SKA should still have the sensitivity to easily resolve sub-L$^{*}$ at high redshifts.  
In the right panel of Figure \ref{fig:tbplots} we illustrate the expected brightness temperature sensitivity of the SKA for resolving sub-L$^{*}$ galaxies forming stars at a rate of $\sim$10\,$M_{\odot}\,$yr$^{-1}$ at  $1 \lesssim z \lesssim 3$.  
For reference, L$^{*}$ galaxies at $1 \lesssim z \lesssim3$ have corresponding total infrared (IR; $8-1000\,\mu$m) luminosities ranging between $10^{11}\,L_{\odot} \lesssim L_{\rm IR} \lesssim 10^{12}\,L_{\odot}$ \citep{2013MNRAS.432...23G}.  
We assume galaxy sizes of 0\farcs65, which is similar to the H$\alpha$ sizes of galaxies in this same redshift range \citep{2013ApJ...763L..16N}.   
Given that star-forming galaxies are likely to host clumpy star formation, the estimated brightness temperature requirements for detecting such galaxies are likely conservative.  
Even so, SKA1-MID will easily resolve such sources at $z\lesssim2$, while also being able to map galaxies at $\lesssim$0\farcs1 resolution that are forming stars at $\sim$100\,$M_{\odot}\,$yr$^{-1}$.   
Such high resolution imaging will be well matched to existing and forthcoming optical/NIR imaging (e.g., JWST) for detailed investigations of resolved star formation and stellar mass.  
Furthermore, resolving highly obscured star-bursting systems at $z\gtrsim1$, with $\approx$500\,pc resolution, will enable critical new investigations of their energetics and distinguish between disks and mergers without being subject to spatially varying obscuration of the rest frame UV light.  
For instance, having angular sizes of each source measured will allow us to calculate brightness temperatures, which can be used to identify maximal starbursts \citep{2005ApJ...618..569M} and AGN \citep{1991ApJ...378...65C} for the brightest sources in the field.    

\subsection{Identifying AGN in the High-Redshift Universe with Band~5}

\subsubsection{Polarization in the Faraday-thin Regime}
At $\nu \gtrsim$10\,GHz, Band~5 observations are well above frequencies at which internal or external Faraday rotation or depolarization effects are important, allowing for a clear view of the intrinsic polarization properties of the sources.  
This is an important complement to polarization observations at lower frequencies.  
By being in the Faraday-thin regime, joint statistical analyses with lower frequencies (e.g., Band 2) will directly measure the depolarization, hopefully shedding light on the outstanding mystery as to whether the faint source population has higher fractional polarization than stronger sources \citep{2002AA...396..463M, 2007ApJ...666..201T} or not \citep{2014MNRAS.440.3113H}.  
Disentangling the role of depolarization and the relation to luminosity and redshift are important keys to this puzzle.  
For example, knowing the statistics of frequency dependence of polarization will let us model the effect of redshift. 

Very broad frequency bands permit modeling for multiple rotation measure components and exploration of the internal physics, magnetic fields and thermal plasma environments of the AGN cores.  
This is something that can only be done at radio wavelengths and provides unique insights into AGN physics on the scales of the accretion disk. 
Polarization information can also be used to sort out cross-identifications of sources; i.e., this information can be used to distinguish radio lobes from completely separate galaxies. 
As already stated, for star-forming galaxies at increasingly high redshift, a significant fraction of the 10\,GHz flux density will be thermal.  
In addition to having high angular resolution imaging for resolving disks versus AGN cores, polarization can play an important role for interpreting the thermal/non-thermal mix of the radiation and separating AGN from disks.
It is only by including Band~5 in SKA1-MID that such polarization studies can be carried out.  




\subsubsection{Maser Lines}
For a more general discussion of maser science that can be done during SKA1 operations, we refer the reader to the chapter by \citep{beswickAASKA14}, and only focus here on what is enabled by the inclusion of Band~5.    
The transition between the 6$_{16}$ -- 5$_{23}$ rotational levels of H$_2$O can show in extragalactic sources as spectacularly strong maser emission at rest frame 22.235 GHz. The maser action results from the collisional excitation of water molecules in extremely dense gas clouds [$n({\rm H}_2) \geq 10^7$ cm$^{-3}$] with temperatures greater than 1000\,K \citep{2005ARAA..43..625L}.
Thus, extragalactic water masers are associated with an AGN, originating in either the circumnuclear accretion disc \citep{1995AA...304...21G} or circumnuclear molecular clouds that are excited by the impact of a jet or AGN outflow \citep{1998ApJ...507L..79C, 2003ApJ...590..149P, 2003ApJ...590..162G}.  
The most luminous of water masers, so called gigamasers, have isotropic line luminosities greater than 10,000 ${L}_\odot$ \citep{2005ApJ...628L..89B, 2008Natur.456..927I}. 
So far only 2 water masers have been discovered at $z > 0.5$, and one of those only because the AGN is lensed, but their discovery suggests the space density of luminous water masers was much higher at high redshift than in the local Universe \citep{2008Natur.456..927I}. 
SKA1-MID will be able to detect gigamaser water emission to $z\sim 2$ in roughly 24\,hr of integration (assuming 4 times the sensitivity of the JVLA), and megamaser emission to $z = 0.5-1$. 

Finding water masers at higher redshifts has the potential to confirm whether the unified model for low luminosity AGN can be extended to the higher luminosity distant sources. 
We can attempt to address whether high QSO luminosities result in ``gigamasers," or whether they destroy the warm dense molecular gas required to supply the water molecules for the maser emission.  
Finding masers in high redshift AGN may provide insight into the size and structure of their circumnuclear molecular clouds or accretion disks, and enable independent measurements of their black-hole masses. Even more importantly, it has the potential to constrain the nature of dark energy through accurate measurement of geometrical distances \citep{1999Natur.400..539H, 2010ApJ...718..657B}, and thus independently verify the results from Type 1A supernova experiments. Measuring a distance with water masers requires the long baselines of the SKA, but SKA1-MID could already make progress in this area with high fidelity very long baseline interferometry (VLBI) imaging.


\section{Looking Towards a Completed SKA \label{sec:ska}}
Restricting SKA1-MID operations to frequencies $<20$\,GHz leaves a gaping hole between where the SKA frequency coverage ends, and ALMA Band 1 begins (i.e., at $\approx$31\,GHz).  
Furthermore, given the technical demands required for achieving the core science goals at $\sim$GHz frequencies (e.g., dynamic range and pointing accuracy), having dishes that are able to operate efficiently at frequencies as high as $\sim$30\,GHz may inevitably become a requirement for SKA1-MID.  
We therefore additionally present some fundamental science goals that are only achievable by keeping the SKA design flexible enough such that future receiver bands at $\gtrsim$20\,GHz could be added to the facility as SKA2 becomes realized.  


The frequency range between 20 and 30\,GHz is currently under-explored in extragalactic astrophysics, largely because the emission from galaxies at these frequencies is intrinsically faint.    
However, this is a critical part of frequency space for proper modeling and accounting of the energetic processes powering galaxies, as this region samples the location where the microwave emission is powered by non-thermal (synchrotron), free-free (bremsstrahlung), and thermal dust emission at somewhat commensurate levels (see Figure \ref{fig:spec}).  
Thus, the ability to properly deconstruct radio spectra into its component parts, which is highly desirable for investigations focusing on the evolution of star-formation because of the need for a clean star formation rate diagnostic  (i.e., free-free continuum), clearly relies on access to this spectral window.  

Similar to the discussion above for observations at $\sim$10\,GHz, in this $20-30$\,GHz band, the sensitivity to free-free emission from a galaxy increases with redshift, as illustrated by the shaded regions in Figure \ref{fig:spec}.  
By redshift $z \sim 2$, where the star formation rate density of the Universe is dominated by extremely dusty starbursts that are essentially opaque at optical/UV wavelengths, a 25\,GHz SKA2 observation samples the rest frame 75\,GHz emission, which is completely dominated by free-free emission arising from young, massive star-forming regions independent of increased synchrotron suppression with redshift due to CMB effects.   
Unlike optical/UV, H$\alpha$, low-frequency radio (synchrotron), and infrared observations, free-free emission in the $\sim30-100$\,GHz range provides a direct measurement of the ionizing photon rate from massive stars independent of dust, making it one of the most robust measurements of a galaxy's star formation rate.    

\subsection{Resolving Individual Star-Forming Regions at $z \gtrsim 1$}
With baselines of 200\,km, observations at 25\,GHz will be able to achieve angular resolutions of $\approx$0\farcs01, probing $\approx$100\,pc scales at $z\gtrsim1$ (i.e., the size of giant molecular clouds and H{\sc ii} regions).   
Given a 1000\,hr exposure, SKA2 should have the brightness temperature sensitivity to resolve such sources having star formation rates of $\sim$100\,$M_{\odot}\,$yr$^{-1}$ with $\approx$0\farcs02 (150\,pc) resolution at $z\sim1$ (right panel of Figure \ref{fig:tbplots}).  
Assuming eventual baselines of $\sim$2000\,km, the array will have the ability to resolve individual star-forming regions on the 10's of pc scales at $z \gtrsim1$, which should be possible for strongly lensed sources.    
Accordingly, SKA2 will enable routine investigations on the distribution of star formation {\it within} galaxies at cosmological distances, when the Universe was forming stars at an order of magnitude higher rate than today.  
Such observations will be ideally complemented by targeted follow-up observations using ALMA to study the molecular gas distribution at similar scales.  
To date, such investigations must rely on chance gravitational lensing to achieve such high spatial resolution \citep[e.g.,][]{2009ApJ...707L.163S, 2011ApJ...742...11S}.  
Consequently, combining these datasets will yield new insight on the physics of star formation by providing a means to investigate whether the relatively constant gas depletion timescale in local galaxy disks persists in the underlying disks of galaxies during the peak of the cosmic star formation rate density at $z\sim2$ \citep[e.g.,][]{2014arXiv1409.8157B}.  
This is a transformational step in the studies of star formation, as presently, such detailed (``high" resolution) investigations of the distributed ($\lesssim$sub-kpc-scale) star formation and molecular gas within galaxies can solely be achieved for objects in the nearby Universe \citep[e.g.,][]{2008AJ....136.2846B, 2011ApJ...730L..13B, 2008AJ....136.2782L, 2013AJ....146...19L}.  
At even higher redshifts, the observed frame 25\,GHz emission will probe the cold dust emission of galaxies, pinning down the Raleigh-Jeans tail of dust spectral energy distributions, which is critical for dust modeling as the cold dust traces the bulk of the dust (and total ISM) mass within galaxies \citep[e.g.,][]{scoville14}.  
By a redshift of $z \sim 5$, such observations with SKA2 would be able to measure rest frame 2\,mm emission.  
And, with only 200\,km baselines, such observations would already be able to measure the cold dust emission on the scales of giant molecular clouds in galaxies just after (and during) the epoch of reionization.

\subsection{Detecting Molecular Gas During the Peak of the Cosmic Star Formation Density}  
In addition to studying the radio-to-infrared continuum at these frequencies, there is a wealth of spectral line studies that can only be achieved through access to this spectral window.  
A detailed discussion of the full potential for molecular gas studies with the SKA is presented in this book by \citet{waggAASKA14}.   
Here we only briefly mention the synergies that can be achieved by combining ALMA and SKA observations for studying molecular gas and star formation at high redshift.  

For investigations of the star formation law \citep{1959ApJ...129..243S, 1998ApJ...498..541K}, which relates the star formation rate and gas surfaces densities in galaxies, a key element is having an estimate of the total molecular gas content, which is well measured by low-$J$ rotational lines of CO; 
higher $J$ transitions, more easily measured with ALMA, trace the warm dense gas and are not as sensitive to the bulk of the molecular gas content in galaxies.    
Both CO ($J=1\rightarrow0$) and CO ($J=2\rightarrow1$) can be measured for galaxies in the redshift range of $z=2.8 - 10.5$ over the full $20-30$\,GHz spectral window (see Figure \ref{fig:spec}).  
Such a capability will have a strong synergy with ALMA, as targeted, high-frequency ALMA observations (e.g., Band 9) will be able to map out the peak of the dust emission in such galaxies (tracing star formation activity), which can be compared to maps of the low-$J$ CO lines obtained using the SKA. 
The joint capability of these two interferometers to determine the full CO spectral line energy distribution of high-$z$ galaxies, from the ground state to high-$J$ transitions, would also facilitate insight into the relative importance of high- and low-excitation components, and hence into the causes for spatial variations of star-formation efficiency within high-$z$ galaxies. While it is to be expected that the decreasing contrast between  line emission and the warmer CMB makes line detections increasingly difficult at $z>4$ \citep{2013ApJ...766...13D}, the SKA is uniquely positioned to detect this low-contrast emission from low rotational levels of molecules at high redshift.
Investigations such as these will become routine, and greatly improve our models of star formation over cosmic time which, to date, are rather unsophisticated in their prescriptions for transforming gas into stars at early times.  

The ability to measure the densest star-forming gas would also be possible via the HCN ($J=1\rightarrow0$) and HNC ($J=1\rightarrow0$) lines, which redshift to frequencies $\lesssim 30$\,GHz for $ z\gtrsim 2$.  
Additionally, the combination of molecular lines and their isotopologues (e.g., $J=1\rightarrow0$ lines of $^{13}$CO and $^{12}$CO) can be used to investigate the physical conditions in galaxies at and beyond the peak of star formation (i.e., $3 \lesssim z \lesssim 5$).  
The ratio of these lines provides a robust measure for the temperature of dense, UV-shielded star-forming cores.  
Such observations can in fact be used to test for initial mass function (IMF) variations \citep[e.g.,][]{2010ApJ...720..226P}, as it is currently thought that a transition to a more top-heavy IMF at early times may be necessary to explain the epoch of reionization \citep[e.g.,][]{chary08}.      
The same can also be said for ratios of free-free to rest-frame UV emission, which are highly sensitive to changes in the IMF.  
There are, of course, radio recombination lines at these frequencies that will also prove useful to redshift identification and for characterizing the conditions of star formation (e.g., radio recombination line-to-continuum ratios can be used to directly measure the electron temperature of H{\sc ii} regions).  

\section{Conclusions}
The SKA is a major scientific investment that will provide the astronomical community with a {\it general use} facility having an expected lifetime of $\gtrsim$50\,yr.   
The inclusion of high-frequency (i.e., $\gtrsim$10\,GHz Band~5) capabilities during SKA1-MID operations, thereby limiting significant redundancy between SKA1-MID and SKA1-SUR capabilities, will yield the highest scientific return by delivering the most flexible telescope to the astronomical community.  

Given these fundamental science goals, including detailed investigations on the evolution and distribution of the star formation and molecular gas content in galaxies into the epoch of reionization, 
it is crucial that the dish design of the SKA be flexible enough to include the possibility of being fit with receivers operating up to 30\,GHz.  
While SKA1-MID should benefit from having some MeerKAT dishes that are fitted with receivers operating near between $8-14.5$\,GHz, and therefore opening up a new window on some of the science topics outlined above,  $20-30$\,GHz capabilities are currently not included in the SKA1 System Baseline Design \citep{ska1_baseline}.  
Furthermore, having contiguous frequency coverage between the SKA and ALMA will ensure that future scientific opportunities, including the ones that we are currently not able to anticipate, will not be missed over the next half century.  

\bibliographystyle{apj}

\end{document}